\documentclass[
amsmath,amssymb,pdftex,
aip, apm, reprint,
]{revtex4-2}

\usepackage{graphicx}
\usepackage{dcolumn}
\usepackage{bm}

\usepackage[utf8]{inputenc}
\usepackage[T1]{fontenc}
\usepackage{mathptmx}
\usepackage{etoolbox}
\usepackage{color}
\usepackage{tabularx}

\makeatletter
\def\@email#1#2{%
 \endgroup
 \patchcmd{\titleblock@produce}
  {\frontmatter@RRAPformat}
  {\frontmatter@RRAPformat{\produce@RRAP{*#1\href{mailto:#2}{#2}}}\frontmatter@RRAPformat}
  {}{}
}%
\makeatother
\begin{document}

\preprint{AIP/123-QED}

\title{Enhanced interlayer electron transfer by surface treatments in mixed-dimensional van der Waals semiconductor heterostructures}
\author{Takeshi~Odagawa}
 \affiliation{Department of Materials Science, Graduate School of Engineering, Tohoku University, 6-6-02 Aramaki-Aza Aoba, Aoba-ku, 980-8579, Sendai, Japan}
\author{Sota~Yamamoto}
 \affiliation{Department of Materials Science, Graduate School of Engineering, Tohoku University, 6-6-02 Aramaki-Aza Aoba, Aoba-ku, 980-8579, Sendai, Japan}
\author{Chaoliang~Zhang}
 \affiliation{Department of Applied Physics, Graduate School of Engineering, Tohoku University, 6-6-05 Aramaki-Aza Aoba, Aoba-ku, 980-8579, Sendai, Japan}
\author{Kazuki~Koyama}
 \affiliation{Department of Materials Science, Graduate School of Engineering, Tohoku University, 6-6-02 Aramaki-Aza Aoba, Aoba-ku, 980-8579, Sendai, Japan}
\author{Jun~Ishihara}
 \affiliation{Department of Materials Science, Graduate School of Engineering, Tohoku University, 6-6-02 Aramaki-Aza Aoba, Aoba-ku, 980-8579, Sendai, Japan}
\author{Giacomo~Mariani}
 \affiliation{NTT Basic Research Laboratories, NTT Corporation, 3-1 Morinosato-Wakamiya, Atsugi, 243-0198, Kanagawa, Japan}
\author{Yoji~Kunihashi}
 \affiliation{NTT Basic Research Laboratories, NTT Corporation, 3-1 Morinosato-Wakamiya, Atsugi, 243-0198, Kanagawa, Japan}
\author{Haruki~Sanada}
 \affiliation{NTT Basic Research Laboratories, NTT Corporation, 3-1 Morinosato-Wakamiya, Atsugi, 243-0198, Kanagawa, Japan}
\author{Junsaku~Nitta}
 \affiliation{Department of Materials Science, Graduate School of Engineering, Tohoku University, 6-6-02 Aramaki-Aza Aoba, Aoba-ku, 980-8579, Sendai, Japan}
 \affiliation{NTT Basic Research Laboratories, NTT Corporation, 3-1 Morinosato-Wakamiya, Atsugi, 243-0198, Kanagawa, Japan}
\author{Makoto~Kohda}
 \affiliation{Department of Materials Science, Graduate School of Engineering, Tohoku University, 6-6-02 Aramaki-Aza Aoba, Aoba-ku, 980-8579, Sendai, Japan}
 \affiliation{Center for Science and Innovation in Spintronics, Tohoku University, 2-1-1 Katahira, Aoba-ku, 980-8577, Sendai, Japan}
 \affiliation{Division for the Establishment of Frontier Sciences of the Organization for Advanced Studies, Tohoku University, 2-1-1 Katahira, Aoba-ku, 980-8577, Sendai, Japan}
 \affiliation{Quantum Materials and Applications Research Center (QUARC), National Institute for Quantum Science and Technology, 1233 Watanuki-Machi, 370-1292, Takasaki, Gunma, Japan}
 \email{makoto.koda.c5@tohoku.ac.jp}

\date{\today}

\begin{abstract}
We investigate the excitonic species in WS$_{2}$ monolayers transferred onto III-V semiconductor substrates with different surface treatments. 
When the III-V substrates were covered with amorphous native oxides, negatively charged excitons dominate the spectral weight in low-temperature near-resonance photoluminescence (PL) measurements. 
However, when the native oxides of the III-V substrates were reduced, neutral excitons begin to dominate the spectral weight, indicating a reduction in the electron density in the WS$_{2}$ monolayers. 
The removal of the native oxides enhanced the electron transfer from the WS$_{2}$ monolayer to the III-V substrate. 
In addition, an additional shoulder-like PL feature appeared $\sim$50 meV below the emission of neutral excitons, which can be attributed to the emission of localized excitons. 
When the III-V substrate surface was passivated by sulfur after the reduction of the native oxides, neutral excitons still dominated the spectral weight. 
However, the low energy PL shoulder disappeared again, suggesting the effective delocalization of excitons through the substrate surface passivation. 
Surface engineering of the semiconductor substrates for two-dimensional (2D) materials can provide a novel approach to control the carrier density of the 2D materials, to implement deterministic carrier localization or delocalization for the 2D materials, and to facilitate the interlayer transfer of charge, spin, and valley currents. 
These findings open the avenue for novel device concepts and phenomena in mixed-dimensional semiconductor heterostructures.
\end{abstract}

\maketitle

\section{\label{introduction}Introduction}
Since the isolation of graphene~\cite{Novoselov16}, a wide variety of two-dimensional (2D) van der Waals (vdW) materials have been studied over the past decades:
a library of 2D vdW crystals has been expanded to semiconducting transition metal dichalcogenides (TMDCs)~\cite{Novoselov05,Mak10,Splendiani10,Manzeli17}, insulating hexagonal boron nitride (hBN)~\cite{Dean10}, and even ferromagnetic CrI$_{3}$~\cite{Huang17} and Cr$_{2}$Ge$_{2}$Te$_{6}$~\cite{Gong17}, as well as ferroelectric and piezoelectric SnTe~\cite{Chang16}, SnS~\cite{Khan20,Higashitarumizu20}, WTe$_{2}$~\cite{Fei18}, and $\alpha$-In$_{2}$Se$_{3}$~\cite{Ding17,Zhou17,Xue18}.
Importantly, such 2D vdW materials generally require a substrate to provide mechanical support, which at the same time significantly influences their properties.
For example, the significant enhancement of the electronic properties of graphene encapsulated by high-quality hBN~\cite{Dean10} emphasizes the importance of the substrate material. 
This means that the interface between 2D materials and substrates which are typically three-dimentional (3D) bulk becomes crucial for harnessing the potential characteristics of 2D materials.

Such 2D/3D heterostructures are promising platform for novel artificial material design.
This is because dangling bond-free surfaces for 2D materials allow the stacking of constituent layers at desired sequences and angles, that can not be realized by epitaxial growth of 2D materials on 3D substrates.
In particular, vdW heterostructures based on 3D III-V semiconductor and 2D TMDC are among promising platforms by taking advantage of high mobility and strong light absorption for III-V semiconductors and spin-valley coupling for TMDCs.
Indeed, variety of state-of-the-art devices have been realized, incluiding a helicity-controllable spin light-emitting diodes~\cite{Ye16}, highly efficient solar cells~\cite{Lin15}, and ultrasensitive photodetectors~\cite{Jia20,Li18}.
For these applications, charge and/or spin transfer between these 2D and 3D materials is critical in the efficiency of device operation.
Interestingly, as will be shown in this paper, such interconnection across the interface is drastically modified only by controlling the surface of 3D materials.
However, the effect of surface treatment itself on both electrical and optical properties of the 2D/3D heterostructures has not been discussed intensively.
Therefore, in this study, we shed light on the surface treatment techniques and comprehensively discuss their effects on the optoelectrical properties of the 2D material.

We focused on monolayer WS$_{2}$/III-V semiconductor heterostructures and demonstrates that reducing the native oxides of the III-V semiconductor enhances the electron transfer from the WS$_2$ to the III-V substrate. 
Furthermore, passivating the reduced III-V substrate by sulfur effectively reduces the spectral weight of localized excitons.


\section{Sample fabrication procedure}

\begin{figure*}
\centering\includegraphics{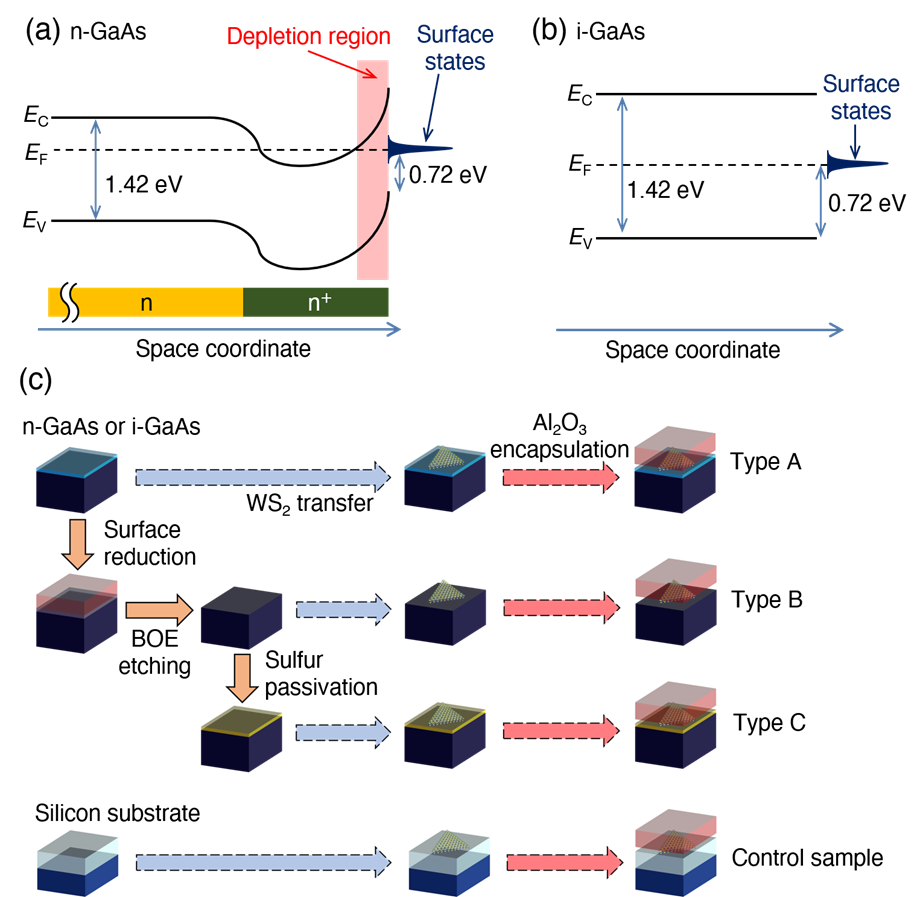}
\caption{
\label{Fig1}
(a) Near-surface band diagram of the $n^+$-In$_{0.04}$Ga$_{0.96}$As and $n$-In$_{0.04}$Ga$_{0.96}$As layers in the $n$-GaAs structure.
A sharp upward band bending in the $n^+$-In$_{0.04}$Ga$_{0.96}$As layer occurs to compensate for the large Fermi level mismatch between the surface and the other deeper region, forming a depletion region.
(b) Near-surface band diagram of the $i$-GaAs structure. 
The $i$-GaAs structure shows no band bending due to the nearly matched Fermi level positions between the surface and the bulk region, in contrast to the $n$-GaAs structure.
(c) Sample fabrication process. 
The native oxides were shown in the light blue layer on the thicker dark blue layer representing the $n$-GaAs and $i$-GaAs structures. 
Type A $n$- and $i$-GaAs samples were degreased with acetone, ethanol, and deionized water (not shown) before transferring CVD-grown WS$_{2}$ monolayer. 
Type B $n$- and $i$-GaAs samples were surface reduced through the self-cleaning effect (surface reduction) after degreasing with acetone, ethanol, and deionized water (not shown), onto which CVD-grown WS$_{2}$ monolayers were transferred. 
The thin gray layer represents the AlO$_{x}$ layer formed from oxygen atoms in the native oxides and aluminum atoms in chemisorbed TMA vapors. 
The thicker red layer on the AlO$_{x}$ layer represents a capping Al$_{2}$O$_{3}$ layer deposited on top of the AlO$_{x}$ layer after the self-cleaning. 
Etching the AlO$_{x}$ and Al$_{2}$O$_{3}$ layers using a BOE exposed the reduced III-V substrate surface. 
Type C $n$- and $i$-GaAs samples were surface reduced through the self-cleaning effect (surface reduction) and sulfur passivated using a CH$_{3}$CSNH$_{2}$/NH$_{4}$OH solution (sulfur passivation) after degreasing with acetone, ethanol, and deionized water (not shown), on which CVD-grown WS$_{2}$ monolayers were transferred. 
The immersion in the CH$_{3}$CSNH$_{2}$/NH$_{4}$OH solution forms a sulfide layer on the III-V substrate surface, passivating the surface dangling bonds.
The control samples were fabricated by transferring CVD-grown WS$_{2}$ monolayers onto a silicon substrate. 
All the samples were encapsulated with an Al$_{2}$O$_{3}$ layer (Al$_{2}$O$_{3}$ encapsulation).
}
\end{figure*}

The III-V semiconductors used in this study were heavily $n$-doped In$_{0.04}$Ga$_{0.96}$As and semi-insulating GaAs structures. 
The $n$-doped structure is composed of a 20 nm $n^+$-In$_{0.04}$Ga$_{0.96}$As layer ($n = 2\times10^{19}$ cm$^{-3}$), a 500 nm $n$-In$_{0.04}$Ga$_{0.96}$As layer ($n = 6\times10^{16}$ cm$^{-3}$) and a 300 nm semi-insulating GaAs buffer layer, epitaxially grown on a (001) semi-insulating GaAs substrate.
The bottom of the conduction band (CB) in the $n^+$-In$_{0.04}$Ga$_{0.96}$As layer is below the Fermi level, while that in the $n$-In$_{0.04}$Ga$_{0.96}$As layer is above the Fermi level.
As shown in Fig.~\ref{Fig1}(a), since the surface Fermi level is at 0.72 eV above the top of the valence band (VB)~\cite{Shen90}, the Fermi level difference between the surface and bulk of the $n^+$-In$_{0.04}$Ga$_{0.96}$As layer induces carrier depletion near the surface, resulting a significant band bending.
On the other hand, the semi-insulating structure consists of nominally undoped (001) GaAs.
As shown in Fig.~\ref{Fig1}(b), the surface Fermi level of this structure is also at 0.72 eV above the top of the VB~\cite{Shen90}. 
Hereafter, the $n$-type and semi-insulating structures are denoted as $n$-GaAs and $i$-GaAs, respectively.

As illustrated in Fig.~\ref{Fig1}(c), the $n$- and $i$-GaAs structures underwent three different surface treatments: Type A, B, and C.
Type A samples were rinsed with acetone, ethanol and, deionized water, leaving their surface covered with amorphous native oxides. 
Type B samples underwent surface reduction through the self-cleaning effect by trimethylaluminum (TMA) vapors during atomic layer deposition (ALD) of Al$_{2}$O$_{3}$~\cite{Hinkle08,Lee09}, after being rinsed with acetone, ethanol, and deionized water. 
The self-cleaning process consisted of 60 cycles of TMA pulse-purge sequences, which effectively removed a significant portion of the native oxides~\cite{Hinkle08,Lee09}.
Subsequently, a buffered oxide etchant (BOE) was used to etch the Al$_{2}$O$_{3}$ layers, leaving the III-V substrates unetched.
Type C samples underwent sulfur passivation by immersion in a CH$_{3}$CSNH$_{2}$/NH$_{4}$OH solution~\cite{Lu96, Petrovykh05}, following degreasing with acetone, ethanol, and deionized water, as well as the above-mentioned surface reduction process.
Highly crystalline WS$_{2}$ monolayers synthesized using a NaCl-assisted chemical vapor deposition (CVD) method were then transferred onto the $n$-GaAs and $i$-GaAs structures using a polycarbonate-based clean transfer method. 
The samples were then encapsulated with an Al$_{2}$O$_{3}$ layer deposited by ALD to prevent physisorption of H$_{2}$O and O$_{2}$ molecules on the WS$_{2}$ surface, which could otherwise destabilize the carrier density in WS$_{2}$~\cite{Tongay13}.

As a control substrate, we used a silicon substrate consisting of thermally oxidized 300 nm SiO$_{2}$/Si where the 300-nm-thick SiO$_{2}$ layer prevents interlayer charge transfer between WS$_{2}$ and Si.

\section{effect of surface treatment on substrate properties}

\begin{figure}
\centering
\includegraphics[width=\hsize]{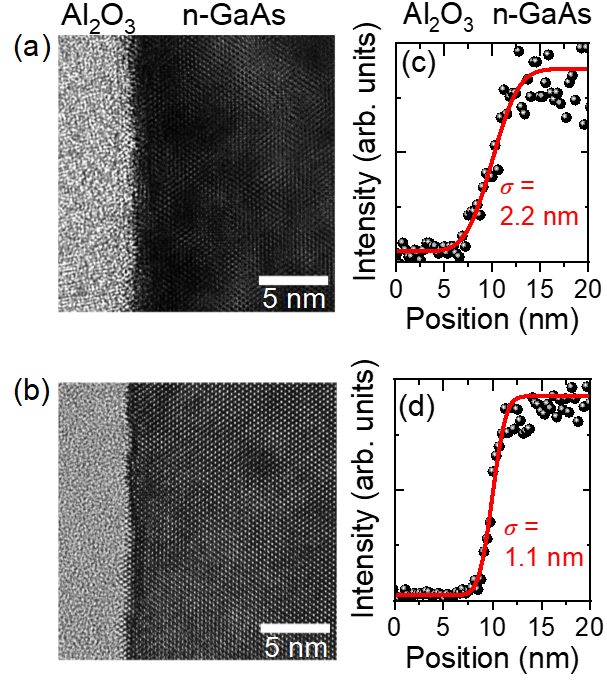}
\caption{
\label{Fig2}
(a), (b) Cross-sectional TEM image of the Al$_{2}$O$_{3}$/$n$-GaAs interface with the single (a) and 60-cycles (b) TMA exposure.
Bright and dark regions are Al$_{2}$O$_{3}$ and $n$-GaAs layers, respectively.
(c), (d) Cross-sectional EDX profile of the Ga and As L shells.
(c) and (d) are respectively for the single and 60-cyclse TMA exposure which are shown in (a) and (b).
The red curves represent fitting curves.
}
\end{figure}

We first confirm that the self-cleaning effect occurs during the Type B and C surface treatments by comparing the samples processed with and without the multiple TMA purge sequences before Al$_{2}$O$_{3}$ deposition.
Figures~\ref{Fig2}(a) and \ref{Fig2}(b) show the transmission electron microscopy (TEM) images of the Al$_{2}$O$_{3}$/$n$-GaAs interfaces with 1 cycle and 60 cycles of TMA pulse-purge secuences, respectively.
Here, while a brighter region corresponds to the Al$_{2}$O$_{3}$ layer, a darker region corresponds to the $n$-GaAs structure.
Since the one pulse-purge sequence reduces only a small fraction of the native oxides~\cite{Lee09}, the interface shown in Fig.~\ref{Fig2}(a) is blurrier than that in Fig.~\ref{Fig2}(b).

For quantitative analysis of the interfacial roughness, we used TEM-coupled energy dispersive X-ray (EDX) spectroscopy to obtain atomic composition profiles for the regions shown in Figs.~\ref{Fig2}(a) and \ref{Fig2}(b).
The EDX line profiles are presented in Figs.~\ref{Fig2}(c) and \ref{Fig2}(d), respectively. 
The sums of the EDX signals from the Ga and As L shells are represented by the black circles. 
The interfacial roughness is extracted using fits to a step-like function $a/2 {\rm erf}[(x - \mu) / (\sqrt{2} \sigma)] + b/2$, shown by the red curves. 
Here, ${\rm erf}(z)$ is the error function with $z$ as a variable, $a$ is step height, $b$ is background signal, $x$ is position, $\mu$ is center position, and $\sigma$ is interfacial roughness.
The interfacial roughness $\sigma$ was improved from 2.2 to 1.1 nm after 60 cycles of TMA pulse-purge sequences, as shown in Fig.~\ref{Fig2}(d).
This decrease in the interfacial roughness indicates the effective removal of the amorphous native oxides, which is consistent with the self-cleaning effect~\cite{Hinkle08, Lee09}.

\begin{figure}
\centering
\includegraphics[width=\hsize]{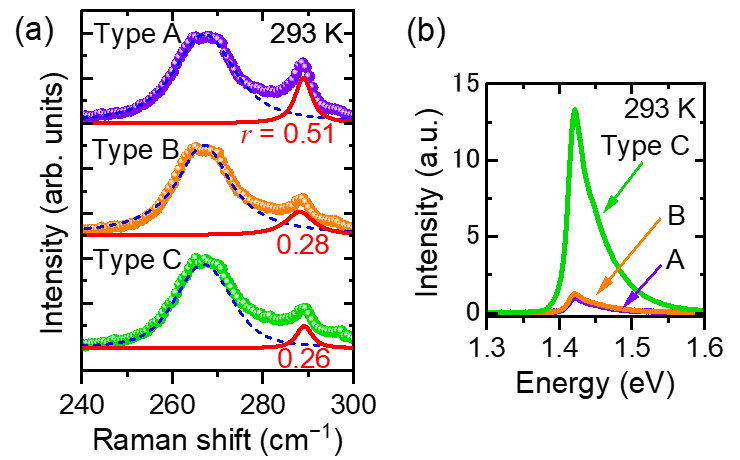}
\caption{
\label{Fig3}
(a) Raman spectra at room temperature for the $n$-GaAs substrate.
$r = I_{{\rm L}^{-}} / I_{\rm LO}$ is the ratio of integral intensity of the L$^{-}$ mode, $I_{{\rm L}^{-}}$, to that of the LO mode, $I_{\rm LO}$.
The LO mode at 288 cm$^{-1}$ and the L$^{-}$ mode at 268 cm$^{-1}$ are deconvolved by a Voight fit.
The fitted LO and L$^{-}$ modes are shown as the red curves and the blue dashed lines, respectively. 
(b) Normalized PL spectra at room temperature for the $i$-GaAs substrate under 2.21 eV excitation.
All the spectra are normalized by the peak PL intensity at 1.42 eV for Type A sample.
}
\end{figure}

The difference due to the surface treatments can be also seen in the Raman spectra measured for the $n$-GaAs structure.
Figure~\ref{Fig3}(a) shows the Raman spectra for the $n$-GaAs structure after transferring WS$_{2}$ monolayer at room temperature in the backscattering configuration under 532 nm excitation. 
The laser penetration depth was estimated to be 124 nm and phonon signal near the $n$-GaAs surfaces  can be observed.
Two distinct peaks at 288 and 268 cm$^{-1}$ are attributed to the longitudinal optical (LO) phonon mode and the lower branch of the LO-phonon-plasmon-coupled (L$^{-}$) mode, respectively~\cite{Yokota61,Varga65}.
The L$^{-}$ mode originates from a un-depleted region and becomes prominent at high carrier density where the coupling between LO phonons and plasmons becomes important~\cite{Yokota61,Varga65,Farrow87}.
Based on its position, the L$^{-}$ mode originates from the un-depleted region in the $n^+$-In$_{0.04}$Ga$_{0.96}$As layer, not in the $n$-In$_{0.04}$Ga$_{0.96}$As layer.
This suggests that charge depletion occurs in the heavily doped $n^+$-In$_{0.04}$Ga$_{0.96}$As layer, allowing us to analyze the depletion region width using the intensity ratio between the LO and L$^{-}$ modes.
It is important to note that Raman scattering from transverse optical phonons, which could spectrally overlap with the LO mode~\cite{Raman23}, is forbidden due to the (001) crystallographic orientation and the backscattering configuration~\cite{Hayes78}.
This selection rule allows for a safe comparison of the intensity ratio $r = I_{{\rm L}^{-}}/I_{\rm LO}$, where $I_{{\rm L}^{-}}$ and $I_{\rm LO}$ represent the integral intensities of the L$^{-}$ and LO modes, respectively.
The integral intensities were determined using a Voigt fit for each peak.
The values of $r$ decrease from 0.51 for Type A to 0.28 and 0.26 for Type B and C, respectively.
This indicates that the self-cleaning and sulfur passivation processes have reduced the thickness of the depletion region.
The reduction is attributed to the decrease in the number of surface states.
When the density of surface states increases, more free electrons are trapped on the surface.

Figure~\ref{Fig3}(b) shows the PL spectra for $i$-GaAs at room temperature under 2.21 eV excitation with a fixed excitation power.
The PL centered at 1.42 eV is attributed to radiative recombination from the CB to the VB~\cite{Tajima88}.
Importantly, the peak intensity of Type B showed a 1.25-fold increase compared to that of Type A.
After the additional sulfur passivation process, Type C showed a substantial 13.3-fold increase of PL intensity compared to Type A.
These facts suggest that non-radiative recombination pathways were reduced due to a decrease in the number of surface states.
These results are consistent with those obtained by Raman spectroscopy~[Fig.~\ref{Fig3}(a)].
It is worth noting that unlike the top layers of the $n$-GaAs structure, the $i$-GaAs structure is nominally undoped and thus the number of free carriers is insufficient to observe the L$^-$ mode.
In the case of the $n$-GaAs structure, the aforementioned sharp band bending repels photoexcited electrons into the bulk region.
Thus, PL intensity is insensitive to the number of surface states.

\section{Effect of surface treatment on heterostructure properties}

Polarization-resolved PL measurements were performed to assign the excitonic species in WS$_{2}$ of the WS$_{2}$/silicon, WS$_{2}$/$n$-GaAs and WS$_{2}$/$i$-GaAs samples.
Both circularly and linearly polarized PL were measured under 2.21 eV excitation at a cryogenic temperature, $T$, of 4 K.
The excitation energy was near-resonant to the exciton PL energy.
In the circularly polarized PL measurements, right-handed ($\sigma_{-}$) and left-handed ($\sigma_{+}$) circularly polarized PL were measured upon photoexcitation of the K valley by $\sigma_{-}$ polarized light, following the spin-valley optical selection rule~\cite{Zeng12,Mak12,Cao12,Jones13,Hsu15,Xiao12}. 
In the linearly polarized PL measurements, on the other hand, horizontal (H) and vertical (V) linearly polarized PL were measured upon photoexcitation of both K and K' valleys by H polarized light.
To ensure reliable analysis of neutral and charged excitons, a combination of circularly and linearly polarized PL measurements was used instead of peak deconvolution.
This approach excludes the effect of peak shifts that may be caused by changes in dielectric screening due to surface treatment of the III-V substrate~\cite{Lin14,Raja17}.

\begin{figure}
\centering
\includegraphics[width=\hsize]{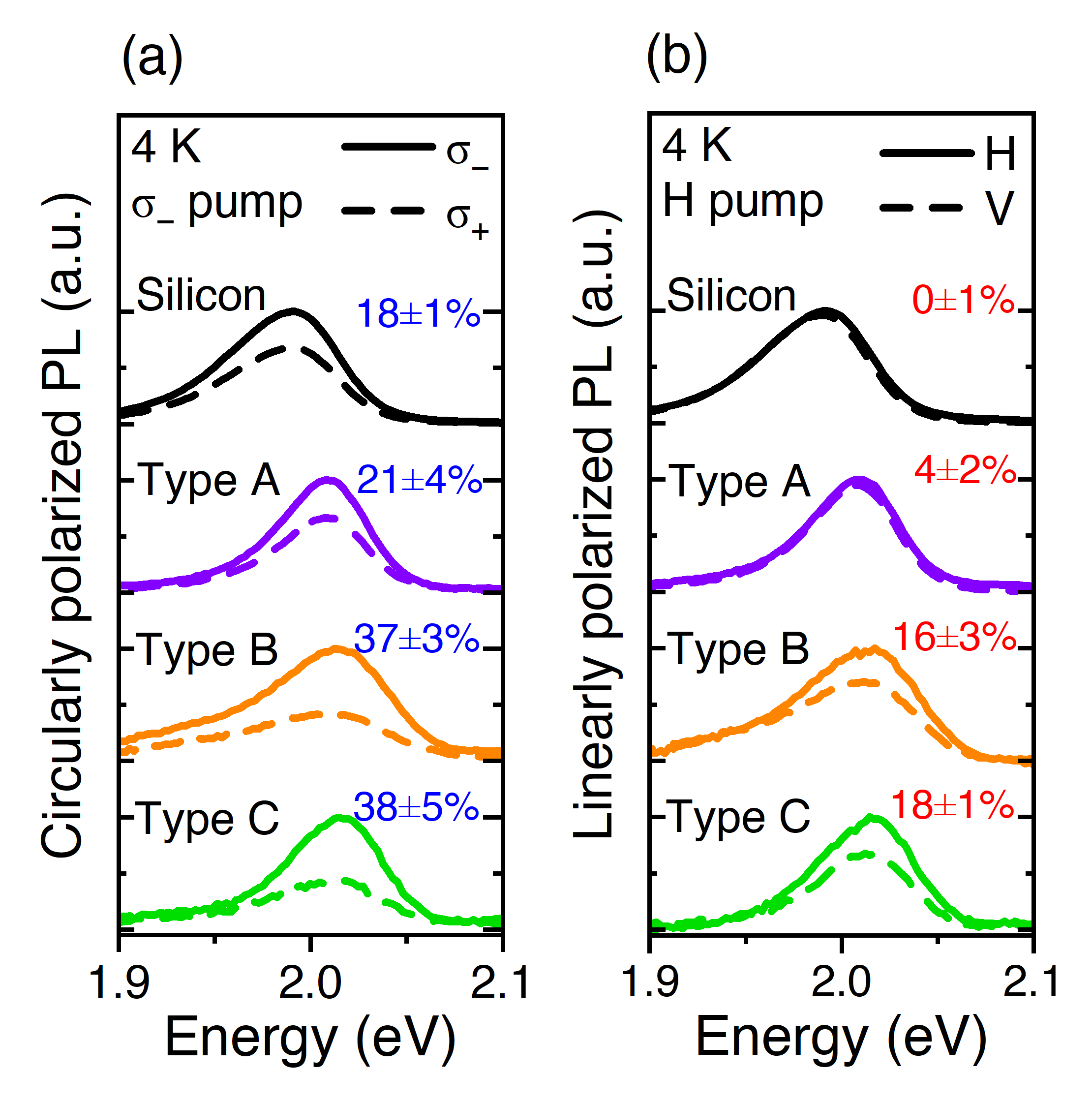}
\caption{
\label{Fig4}
Polarization-resolved PL spectra for the WS$_{2}$ monolayers on silicon and $n$-GaAs substrates.
(a) Circularly polarized PL spectra. 
The degree of circular polarization $\rho_{\rm c}$ for each sample is shown in blue.
(b) Linearly polarized PL spectra. 
The degree of linear polarization $\rho_{\rm l}$ for each sample is shown in red.
}
\end{figure}

Figure~\ref{Fig4}(a) shows representative circularly polarized PL spectra for the WS$_{2}$/silicon (black) and the WS$_{2}$/$n$-GaAs with Type A, B, and C treatments (purple, orange, and green).
In each pair of spectra, the solid and dashed curves represent $\sigma_-$ and $\sigma_+$ polarized PL spectra, respectively.
Degree of circular polarization $\rho_{\rm c}=(I_{-} - I_{+}) / (I_{-} + I_{+})$ for each sample is summarized in Table~\ref{Tab1} where $I_{\mp}$ is the intensity of $\sigma_{\mp}$ polarized PL averaged within $\pm 20$ meV around the peak energy.
The large $\rho_{\rm c}$ independent of the substrate conditions means all samples realized excitonic valley polarization~\cite{Zeng12,Mak12,Cao12,Jones13,Hsu15,Xiao12}.
In other words, depolarization pathways through the electron-hole (e-h) Coulomb exchange interaction~\cite{Xiao12,Yu14} and simultaneous intervalley scattering of electrons and holes by two longitudinal acoustic (LA) phonons~\cite{Asakura21,Kioseoglou16,Tornatzky18} were effectively suppressed. 
This is indeed plausible because the broken inversion symmetry $E_{\uparrow}(k) \neq E_{\uparrow}(-k)$, where $k$ represents the crystal momentum, should still persist.
This is in contrast to the case of bilayer TMDCs~\cite{Mak12}.

\begin{table}
\centering
\caption{\label{Tab1}Summary of $\rho_{\rm c}$ and $\rho_{\rm l}$ observed in WS$_{2}$ on $n$-GaAs.}
\begin{tabularx}{\hsize}{cXcXc}
\hline \hline
Substrate & & $\rho_{\rm c}$ & & $\rho_{\rm l}$ \\
\hline
Silicon & & $18 \pm 1\%$ & & $0 \pm 1\%$ \\
Type A & & $21 \pm 4\%$ & & $4 \pm 2\%$ \\
Type B & & $37 \pm 3\%$ & & $16 \pm 3\%$ \\
Type C & & $38 \pm 5\%$ & & $18 \pm 1\%$ \\
\hline\hline
\end{tabularx}
\end{table}

%
%


Figure~\ref{Fig4}(b) shows linearly polarized PL spectra for the WS$_{2}$/silicon (black) and the WS$_{2}$/$n$-GaAs with Type A, B, and C treatments (purple, orange, and green) measured at the same spot as in Fig.~\ref{Fig4}(a).
In each pair of spectra, the solid and dashed curves represent H and V polarized PL spectra, respectively.
Degree of linear polarization $\rho_{\rm l}=(I_{\rm H} - I_{\rm V})/(I_{\rm H} + I_{\rm V})$ for each sample is summarized in Table~\ref{Tab1} where $I_{\rm H(V)}$ is the intensity of H(V) polarized PL averaged within $\pm 20$ meV around the peak energy.
By analyzing changes in $\rho_{\rm l}$, we assign the dominant excitonic species that contribute the observed PL lines as follows.

Previous studies have shown that in low carrier density, the PL spectrum of WS$_{2}$ is dominated by a Coulomb-bound electron-hole pair, known as a neutral exciton (X$^0$)~\cite{Jones13,Hsu15}.
At $T = 4$ K, where our measurements were performed, PL emission from X$^{0}$ under linearly polarized excitation is considered to show $\rho_{\rm l} \neq 0$ as follows.
A linear superposition of K and K' valleys created by linearly polarized light emits the same linearly polarized light if the recombination lifetime is shorter than the valley decoherence time.
In our case, this situation is plausible since the valley depolarization was effectively suppressed untill recombination as confirmed from $\rho_{\rm c} \neq 0$ under the circularly polarized excitation [Fig.~\ref{Fig4}(a)].
In higher carrier density, on the other hand, X$^{0}$ captures the excess electron or hole, and forms negatively or positively charged excitons (X$^{-}$ or X$^{+}$), respectively.
The PL from both X$^{-}$ and X$^{+}$ are not linearly polarized (i.e., $\rho_{\rm l} = 0$) even if valley depolarization is suppressed.
This is because the optical transition is allowed for circularly polarized light rather than linearly polarized light due to the eigenstate mixing of excitonic state induced by the exchange interaction~\cite{Jones13,Xiao12}.
Therefore, a larger $\rho_{\rm l}$ indicates that X$^{0}$ is dominant.

For the WS$_{2}$/silicon samples, X$^{-}$ or X$^{+}$ dominates the spectral weight since $\rho_{\rm l} = 0 \pm 1\%$ was observed.
The small $\rho_{\rm c}$ values agree well with the $n$-type behavior of CVD-grown WS$_{2}$ monolayers due to the energetically preferred formation of sulfur vacancies~\cite{Amani15}.
Therefore, we assign X$^{-}$ as the dominant excitonic species in the WS$_{2}$/silicon samples.

For the Type A WS$_{2}$/$n$-GaAs samples, relatively small $\rho_{\rm l} = 4 \pm 3\%$ suggests that either X$^{-}$ or X$^{+}$ dominates the spectral weight.
In addition to the fact that the CVD-grown WS$_{2}$ monolayers behave $n$-type as mentioned above~\cite{Amani15}, the energy gap of native oxides $\sim$4 eV~\cite{Mikoushkin18} is large enough to disturb the electron transfer from the WS$_{2}$ to the substrate even after the heterostructure fabrication.
Furthermore, the heavily $n$-doped substrate cannot provide sufficient holes to WS$_{2}$.
Therefore, X$^{-}$ still dominates the spectrum from Type A samples similar to the WS$_{2}$/silicon samples.

On the other hand, the Type B and C WS$_{2}$/$n$-GaAs samples exhibited relatively large $\rho_{\rm l}$, $16\pm3\%$ and $18 \pm 1\%$, respectively.
This indicates X$^{0}$ emission increased, suggesting a reduction in electron density compared to the cases of the WS$_{2}$/silicon and Type A WS$_{2}$/$n$-GaAs samples. 
Therefore the self-cleaning effect due to the TMA pulse-purge sequences successfully removed the native oxides on the substrate surface and thus lowered the energy barrier between WS$_{2}$ and the substrate.

The impact of sulfur passivation can be seen in a PL shoulder at around 1.95 eV that is present in Type B but absent in Type A and C~[Figs.~\ref{Fig4}(a) and \ref{Fig4}(b)].
This shoulder was reproduced in all Type B samples measured, not just the Type B sample shown in Fig.~\ref{Fig4}.
As has been reported in MoS$_{2}$/GaAs heterostructures~\cite{Rojas-Lopez21}, the PL shoulder is considered to originate from exciton localization or interfacial charge trap.
The presence of the PL shoulder in the WS$_{2}$/Type B $n$-GaAs samples and the absence of such shoulder in the WS$_{2}$/Type C $n$-GaAs samples suggest that sulfur passivation plays a significant role in reducing exciton localization or interfacial charge trap in monolayer WS$_{2}$.
This is consistent with the broadest LO peak for Type B $n$-GaAs in Fig.~\ref{Fig3}(a) since the dangling bonds can be the origin of surface states which is in principle disordered and leads to broadening of Raman peak.
Consequently, the use of the III-V substrates with well-controlled surface states plays a significant role for both localization and delocalization of electrons in monolayer WS$_{2}$/III-V semiconductor heterostructures.

The underlying mechanism behind the change in the excitonic species is discussed below. 
The topmost $n^{+}$-In$_{0.04}$Ga$_{0.96}$As layer of the $n$-GaAs structure is $n$-doped and has a thin depletion region.
The modification in both surface states and depletion regions may be dominant factors.
To investigate these mechanisms, the semi-insulating $i$-GaAs structures were used to exclude the effect of the depletion regions. 

%

\begin{figure}
\centering
\includegraphics[width=8.5cm]{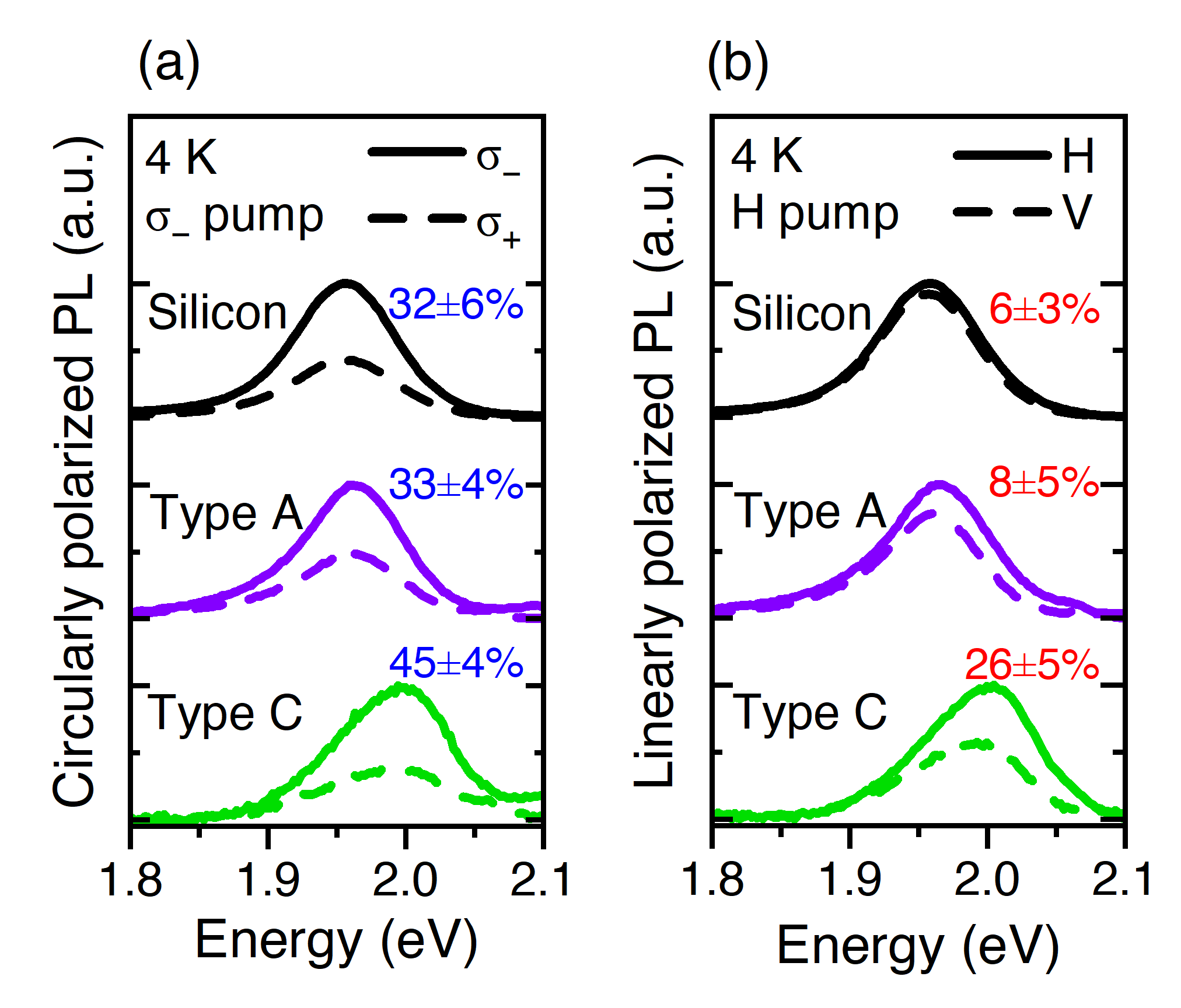}
\caption{
\label{Fig5}
Polarization-resolved PL spectra for the WS$_{2}$ monolayers of the WS$_{2}$/silicon and WS$_{2}$/$i$-GaAs samples. 
(a) Circularly polarized PL spectra.
The degree of circular polarization $\rho_{\rm c}$ for each sample is shown in blue.
(b) Linearly polarized PL spectra. 
The degree of linear polarization $\rho_{\rm l}$ for each sample is shown in red.
}
\end{figure}

\begin{table}
\centering
\caption{\label{Tab2}Summary of $\rho_{\rm c}$ and $\rho_{\rm l}$ observed in WS$_{2}$ on $i$-GaAs.}
\begin{tabularx}{\hsize}{cXcXc}
\hline \hline
Substrate & & $\rho_{\rm c}$ & & $\rho_{\rm l}$ \\
\hline
Silicon & & $32 \pm 6\%$ & & $6 \pm 3\%$ \\
Type A & & $33 \pm 4\%$ & & $8 \pm 5\%$ \\
Type B & & NA & & NA \\
Type C & & $45 \pm 4\%$ & & $26\pm5\%$ \\
\hline\hline
\end{tabularx}
\end{table}

Figure~\ref{Fig5}(a) shows representative circularly polarized PL spectra for the WS$_{2}$/silicon and WS$_{2}$/$i$-GaAs samples with different surface treatments.
All samples show a clear difference between the $\sigma_{-}$ and $\sigma_{+}$ components, indicating effective suppression of the e-h Coulomb exchange~\cite{Xiao12,Yu14} and LA phonon~\cite{Asakura21,Kioseoglou16,Tornatzky18} depolarization pathways. 
Similar to the $n$-GaAs cases, this fact allows to identify the excitonic species through the degree of linear polarization.

Figure~\ref{Fig5}(b) shows the linearly polarized PL spectra measured at the same spot as in Fig.~\ref{Fig5}(a) where H (V) polarized PL are indicated as solid (dashed) curves.
As summarized in Table~\ref{Tab2}, $\rho_{\rm l}$ is larger in the Type C WS$_{2}$/$i$-GaAs samples (green) compared to the WS$_{2}$/silicon (black) and Type A WS$_{2}$/$i$-GaAs samples (purple).
This suggests a lower electron density in the Type C WS$_{2}$/$i$-GaAs samples compared to the WS$_{2}$/silicon and Type A WS$_{2}$/$i$-GaAs samples, similar to the $n$-GaAs case. 
%
It should be noted that the batches used to synthesize WS$_{2}$ were different in Figs.~\ref{Fig4} and \ref{Fig5}.
This must lead to the batch-to-batch difference in the quality of the WS$_{2}$ monolayers, including defect and carrier density, resulting in the difference in PL polarization degrees, for example, as seen in WS$_{2}$/silicon samples.
However, quantitative comparison within the same figure should make sense.

Regardless of the substrate doping concentrations, the Type A samples exhibit lower values of $\rho_{\rm l}$, i.e., a dominance of X$^{-}$ over X$^{0}$.
In contrast, the Type B and C samples exhibit an increased spectral weight of X$^{0}$.
Namely, the reduction in the electron density in WS$_{2}$ is observed regardless of the presence or absence of depletion regions near the surface of the III-V semiconductors. 
This suggests that the space charges in the depletion regions do not play a dominant role in reducing the electron density in WS$_{2}$.
However, the Type B and C WS$_{2}$/$n$-GaAs samples exhibit similar $\rho_{\rm l}$ regardless of the reduction of surface states. 
This suggests that the space charges at the surface states do not play a significant role.
Because the space charges in the depletion regions and at the surface do not play a role for changing the carrier density in monolayer WS$_{2}$, we conclude that the electron transfer from the CB of monolayer WS$_{2}$ to that of III-V semiconductor plays a role in reducing the carrier density in the WS$_{2}$/Type B $n$-GaAs and WS$_{2}$/Type C $n$-GaAs samples.

\section{conclusion}

We focused on WS$_{2}$/$n$-GaAs and WS$_{2}$/$i$-GaAs heterostructures and investigated the effect of surface treatments for the III-V substrates on the excitonic species in the WS$_{2}$ monolayer. 
The PL emission from the WS$_{2}$ monolayers was dominated by the X$^{-}$ feature in the WS$_{2}$/untreated $n$-GaAs and WS$_{2}$/untreated $i$-GaAs samples, regardless of the substrate doping concentrations.
However, in the case of the WS$_{2}$/surface-reduced $n$-GaAs, WS$_{2}$/sulfur-passivated $n$-GaAs and WS$_{2}$/sulfur-passivated $i$-GaAs samples, the X$^{0}$ feature dominated the PL spectra, indicating a reduced electron density in the WS$_{2}$ monolayers. 
The removal of the native oxides may enhance the electron transfer from the WS$_{2}$ monolayer to the III-V substrates.
This means that the relatively low spin injection efficiency, for example, from (Ga,Mn)As to WS$_{2}$ monolayer can be improved by removing the native oxides of the (Ga,Mn)As substrate~\cite{Ye16}.
Furthermore, for the WS$_{2}$/surface-reduced $n$-GaAs samples, an additional PL shoulder appears at $\sim$50 meV lower than the X$^0$ feature, suggesting exciton localization or interfacial charge trap.
However, such a PL shoulder was absent in the WS$_{2}$/sulfur-passivated $n$-GaAs samples.
Therefore, the surface passivation of the III-V substrate may play an important role in suppressing exciton localization or interfacial charge trap in the WS$_{2}$ monolayer, which can lead to high mobility devices~\cite{Dean10}.
These findings could lead to novel device concepts, including gate-free doping and deterministic carrier localization or delocalization for 2D materials.

\section*{Supplementary Material}
The supplementary material contains detailed procedures for the surface processing, the WS$_{2}$ growth and transfer, and the PL measurements.

\begin{acknowledgments}
T.O. and K.K. acknowledge the financial support from the Graduate Program in Spintronics (GP-Spin) at Tohoku University. 
T.O. was financially supported by the JST-SPRING program  (Grant No. JPMJSP2114). 
T.O. acknowledges the technological support from Dr. Kosei Kobayashi at the Graduate School of Engineering, Tohoku University. 
This work was supported by the JSPS KAKENHI program (Grant No. 21H04647) and the JST-FOREST and CREST programs (Grant Nos. JPMJFR203C and JPMJCR22C2).
\end{acknowledgments}

\section*{AUTHOR DECLARATIONS}
\subsection*{Conflict of Interest}
The authors have no conflicts to disclose.

\subsection*{Author Contributions}
\textbf{Takeshi~Odagawa}: Data curation (lead); Formal analysis (lead); Investigation (lead); Methodology (equal); Resources (lead); Validation (lead); Visualization (lead); Writing - original draft (lead); Writing - review \& editing (equal).
\textbf{Sota~Yamamoto}: Formal analysis (equal); Validation (equal); Writing - review \& editing (equal).
\textbf{Chaoliang~Zhang}: Methodology (equal); Resources (equal); Formal analysis (equal); Validation (equal); Writing - review \& editing (equal).
\textbf{Kazuki~Koyama}: Resources (equal); Validation (equal); Writing - review \& editing (equal).
\textbf{Jun~Ishihara}: Formal analysis (equal); Validation (equal); Writing - review \& editing (equal).
\textbf{Giacomo~Mariani}: Validation (equal); Writing - review \& editing (equal).	 
\textbf{Yoji~Kunihashi}:  Validation (equal); Writing - review \& editing (equal).
\textbf{Haruki~Sanada}: Validation (equal); Writing - review \& editing (equal).
\textbf{Junsaku~Nitta}: Conceptualization (equal); Project administration (equal); Writing - review \& editing (equal).
\textbf{Makoto~Kohda}: Conceptualization (lead); Methodology (lead); Project administration (lead); Writing - review \& editing (lead).

\section*{Data Availability Statement}

The data that support the findings of this study are available from the corresponding author upon reasonable request.

\appendix

\bibliography{reference}

\end{document}